\PassOptionsToPackage{pdfpagelabels=false}{hyperref} 
\documentclass[fleqn]{mnras}

\usepackage{amsmath}

\usepackage{natbib}
\usepackage{textcomp}
\usepackage{gensymb}
\usepackage{graphicx}
\usepackage[utf8x]{inputenc}
\usepackage{appendix}

\usepackage{rotating} 
\usepackage{xcolor}
\usepackage[export]{adjustbox}

\definecolor{pink}{rgb}{0.55,0,0.52}
\definecolor{mygreen}{rgb}{0.19,0.55,0.11}
\definecolor{dkgreen}{rgb}{0,0.6,0}
\definecolor{gray}{rgb}{0.5,0.5,0.5}
\definecolor{mauve}{rgb}{0.58,0,0.82}
\definecolor{verbgray}{gray}{0.9}
\definecolor{lightblue}{rgb}{0.85,0.9,1}
\definecolor{lightgreen}{rgb}{0.85,1,0.85}
\definecolor{lightorange}{rgb}{1,0.94,0.8}
\definecolor{forestgreen}{rgb}{0.1,0.49,0.07}

\graphicspath{{./}{Figures/}}

\title[Kinematics of radio galaxies in galaxy groups]{Radio galaxies in galaxy groups: kinematics, scaling relations and AGN feedback}

\author[T. Pasini, A. Finoguenov, M. Brüggen et al.]{
\Large
T. Pasini$^{1}$\thanks{E-mail: thomas.pasini@hs.uni-hamburg.de},
A. Finoguenov$^{2}$,
M. Brüggen$^{1}$,
M. Gaspari$^{3,4}$,
F. de Gasperin$^{1,5}$
and G. Gozaliasl$^{6,2}$\\
$^{1}$Hamburger Sternwarte, Universität Hamburg, Gojenbergsweg 112, 21029 Hamburg, Germany\\
$^{2}$Department of Physics, University of Helsinki, P.O. Box 64, FI-00014 Helsinki, Finland\\
$^{3}$INAF - Osservatorio di Astrofisica e Scienza dello Spazio, via P. Gobetti 93/3, I-40129 Bologna, Italy\\
$^{4}$Department of Astrophysical Sciences, Princeton University, 4 Ivy Lane, Princeton, NJ 08544-1001, USA\\
$^{5}$INAF - Istituto di Radioastronomia, via P. Gobetti 101, 40129, Bologna, Italy\\
$^{6}$Research Program in Systems Oncology (Oncosys), Faculty of Medicine, University of Helsinki, P.O.Box 63, FI-00014, Helsinki, Finland}

\date{Accepted 2021 May 17. Received 2021 May 17; in original form 2021 January 20}

\begin{document}

\pagerange{\pageref{firstpage}--\pageref{lastpage}}
\maketitle
\label{firstpage}

\begin{abstract}
We investigate the kinematic properties of a large (N=998) sample of COSMOS spectroscopic galaxy members distributed among 79 groups. We identify the Brightest Group Galaxies (BGGs) and cross-match our data with the VLA-COSMOS Deep survey at 1.4 GHz, classifying our parent sample into radio/non-radio BGGs and radio/non-radio satellites. The radio luminosity distribution spans from $L_R\sim2\times10^{21}$ W Hz$^{-1}$ to $L_R\sim3\times$10$^{25}$ W Hz$^{-1}$. A phase-space analysis, performed by comparing the velocity ratio (line-of-sight velocity divided by the group velocity dispersion) with the galaxy-group centre offset, reveals that BGGs (radio and non-radio) are mostly ($\sim$80\%) ancient infallers.
Furthermore, the strongest ($L_R>10^{23}$ W Hz$^{-1}$) radio galaxies are always found within 0.2$R_{\rm vir}$ from the group centre. Comparing our samples with HORIZON-AGN, we find that the velocities and offsets of simulated galaxies are more similar to radio BGGs than to non-radio BGGs, albeit statistical tests still highlight significant differences between simulated and real objects. We find that radio BGGs are more likely to be hosted in high-mass groups. Finally, we observe correlations between the powers of BGG radio galaxies and the X-ray temperatures, $T_{\rm x}$, and X-ray luminosities, $L_{\rm x}$, of the host groups. This supports the existence of a link between the intragroup medium and the central radio source. The occurrence of powerful radio galaxies at group centres can be explained by Chaotic Cold Accretion, as the AGN can feed from both the galactic and intragroup condensation, leading to the observed positive $L_{\rm R}-T_{\rm x}$ correlation.
\end{abstract}
\begin{keywords}
galaxies: clusters: general -- galaxies: clusters: intracluster medium -- galaxies: groups: general -- X-rays: galaxies: clusters -- radio continuum: galaxies
\end{keywords}  

\section{Introduction}
\label{sec:intro}

The hot plasma inside of galaxy clusters and groups is governed by processes that can be observed at multiple wavelengths: from thermal cooling of the hot ($\sim$ 10$^7$ K) intra-cluster medium (ICM) \citep[e.g.,][]{Fabian_1994, Peterson-Fabian_2006}, to line emission produced by warm gas \citep[e.g.,][]{Hamer_2016, Pulido_2018}, to feedback from central Active Galactic Nuclei (AGN). The latter can heat their surroundings and can prevent the catastrophic cooling of the cool core \citep[see reviews by, e.g.,][]{McNamara-Nulsen_2007, Gitti_2012}, establishing what is known as \textit{AGN feedback cycle}.  

Brightest Cluster Galaxies (BCGs) and Brightest Group Galaxies (BGGs) are the most optically luminous and massive galaxies in a cluster and group, respectively. Usually they lie at the centres of their host structures. Owing to their special location, the evolution and assembly history of massive galaxies and of their hosts has been studied widely \citep{Bernstein_2001, Bernardi_2007, Liu_2009, Stott_2010}. Notably, BGGs look dissimilar from other massive galaxies, showing different surface brightness profiles and obeying different scaling relations, which suggests that their formation process may be different, too \citep[see e.g.,][]{vonderLinden_2007, Liu_2008, Stott_2008, Shen_2014}. 

In galaxy clusters, the evolution of BCGs is tightly linked to that of the host cluster \citep[e.g.][]{Lin_2007}. 
Several observational studies \citep[e.g.,][]{Giodini_2010, Giacintucci_2011b, Ineson_2013, Ineson_2015, Kolokythas_2018, Pasini_2020} and numerical simulations (\citealt{Gaspari_2020} for a review) have demonstrated the importance of AGN feedback in galaxy groups, that are known to be the hosts of more than half of all galaxies \citep{Eke_2006}. However, there is no consensus on where to draw the boundary between galaxy clusters and galaxy groups. It is sometimes assumed that the boundary lies close to the virial temperature of 1 keV since this is the temperature where the slope of the relation between X-ray luminosity and virial temperature changes. However, this change of slope could be caused by observational systematics \citep[see e.g.,][]{Voit_2018}. 

The trigger mechanism for radio-loud AGN activity is still unclear 
\citep[e.g.,][]{Shakura_1973, Merloni_2007, Best_2012}, but gas that cools out of the hot X-ray halo seems to play a key role (e.g., \citealt{Best_2005}).
In the \textit{Chaotic Cold Accretion} (CCA) scenario \citep[][]{Gaspari_2013,Gaspari_2016} the AGN is frequently switched on and off through self-regulated feeding and feedback cycles. CCA can occur in every galaxy with a hot halo, regardless of the position. However, central galaxies - BGGs or not - lie in dense regions, where the condensation is significantly stronger \citep{Gaspari_2019}. Non-linear thermal instabilities produced by the cooling plasma lead to precipitation, which is able to feed the SMBH through inelastic collisions between the condensed cold clouds and filaments (e.g., \citealt{Gaspari_2017,McDonald_2018,Tremblay_2018,Temi_2018,Juranova_2019,Rose_2019,Schellenberger_2020}).

\citet{Pasini_2020} presented a study of the relation between the ICM X-ray luminosity of a sample of 247 X-ray selected galaxy groups in COSMOS \citep{Gozaliasl_2019} and the radio luminosity produced by the corresponding central radio galaxy, defined as the radio source at 1.4 GHz found closest to the X-ray emission peak. Cross-matching this sample with optical catalogs, they found that only in 30 per cent of the groups central radio galaxies were hosted in BGGs. This is consistent with \citet{Gozaliasl_2019}, who showed that 70 per cent of COSMOS BGGs are found more than $0.1\,R_{200}$ away from the X-ray peak. This suggests that BGGs do not always lie at the bottom of the potential well. This does not seem to be the case for galaxy clusters, where $\sim$ 85 per cent of central radio galaxies were found in BCGs \citep{Pasini_2020}. Nevertheless, there are some cases in which an apparently brightest galaxy near a cluster centre has a significantly large velocity offset with respect to the mean redshift of cluster members ($>300\,\mathrm{km}\,{{\rm{s}}}^{-1}$; \citealt{Coziol_2009, Lauer_2014}), suggesting that these objects may not reside at the bottom of the cluster potential well.  

Recent work \citep[e.g.,][]{Rhee_2017, Gozaliasl_2020} combines the cluster-centric velocities and cluster-centric radii in a single diagram. This phase-space diagram can be used to extract information about the assembly history of clusters \citep[e.g.]{2011MNRAS.416.2882M, 2014MNRAS.438.2186H}. For example, one expects recently accreted galaxies to show higher relative velocities and offsets from the centre than objects accreted at an earlier time. Objects accreted early are usually found within the core of the virialised region and show a small velocity spread \citep{Noble_2016, Gozaliasl_2020}.

In this paper, we investigate the kinematics of the hosts of radio galaxies in groups (BGGs and 'satellites\footnote{Throughout this work, we will refer to non-BGGs as satellites for easier reading}'), comparing them to the kinematics of galaxies with no detected radio emission. To this end, we rely on a recently published sample of X-ray galaxy groups \citep{Gozaliasl_2019}, combining it with optical, kinematic and spectroscopic data. All host groups were identified in the 2 square degree COSMOS field, with a mass range of $M_{200} = 8\times 10^{12}-3\times 10^{14} M_\odot$, where the upper limit of this range corresponds to a virial temperature of $\sim 4$ keV.

This paper is structured as follows: In Sec.~2 we describe our sample and how we compiled it. In Sec.~3.1 we perform a phase-space analysis and in Sec.~3.2, we compare it to cosmological simulations. In Sec.~3.3 we explore the properties of our sample and derive scaling relations in Sec.~3.4. In Sec.~3.5 we discuss implications for AGN feedback before we conclude in Sec.~4.

Throughout the paper, we assume a standard $\Lambda$CDM cosmology with H$_0 = 70$ km s$^{-1}$ Mpc$^{-1}$, $\Omega_\Lambda = 0.73$ and $\Omega_{\text{M}} =  1-\Omega_\Lambda  = 0.27$.

\section{The sample}

\citet{Gozaliasl_2019} presented a sample of 247 X-ray selected galaxy groups in the 2 square degree COSMOS field at a redshift range of $0.08 \leq z < 1.53$. The same sample was also studied in \citet{Pasini_2020} making use of radio data from the VLA-COSMOS survey \citep{Schinnerer_2010} and of new MeerKAT observations that are part of the MIGHTEE survey \citep{Jarvis_2016}. In this paper, they found evidence for a correlation between the X-ray luminosity of galaxy groups and the radio luminosity of the central AGN (see \citealt{Pasini_2020} for further details).

Before we can perform a dynamical analysis, we need to determine cluster membership through spectroscopic redshifts. To this end, we have vetted the group membership catalog of \citet{Gozaliasl_2019} by applying the {\sc Clean} algorithm of \cite{Mamon_2013}, which removes the galaxies exceeding the escape velocity of the group as a function of cluster-centric radius from the group. We kept groups with more than four member galaxies and removed all of those galaxies that have no spectroscopic data. This resulted in a total of 79 groups, with 998 member galaxies, which limits our study to redshifts of $z \leq 1.0$. Among these members, we identified 70 BGGs, that were found to be the most massive in each group by \citet{Gozaliasl_2014, Gozaliasl_2019}. We have computed the gapper velocity dispersion estimates $\sigma_\mathrm{v}$ following \citet{Beers90}.

If the most massive galaxy only has photometric redshifts, it gets excluded from our sample. As a result, not every group has its BGG included in this work, and for some groups the BGG identification could even be wrong if another galaxy was mistakenly identified as BGG. Therefore, we performed a further check and found this to be the case only in one group, where the galaxy previously classified as BGG was actually a satellite. 

The spectroscopic member galaxies thus obtained were then matched to the VLA-COSMOS Deep survey at 1.4 GHz (rms $\sim$ 12 $\mu$Jy beam$^{-1}$, beam = 2.5\arcsec x 2.5\arcsec, \citealt{Schinnerer_2010}).
The cross-match was performed by assuming an association between radio emission and optical galaxy when their angular distance is less than the width of the beam in the VLA-COSMOS survey. We found that 50 groups host at least 1 radio source according to our criteria, with a total of 79 detected radio galaxies (28 in BGGs and 51 in satellites). 19 of these groups host more than one radio galaxy, while for 31 we only detect one source. Groups with no detected radio emission are therefore 29, with 236 member galaxies in total (23 BGGs and 213 satellites). The remaining 683 galaxies (19 BGGs and 664 satellites) with no radio emission belong to the 50 groups with at least one radio galaxy. The characteristics of the samples are briefly summarized in Table \ref{tab:properties}.

\begin{table}
	\centering 
	\begin{tabular}{c c c}
		\hline
		\hline
		& BGGs & Satellites \\
		\hline
		Radio detection & 28 & 51 \\
		No radio detection & 42$^a$ & 877$^b$ \\
		\hline
		Total & 70 & 928 \\
		\hline
	\end{tabular}
	\caption{Properties of the samples studied in this work. \\ $^a$: 19 from the 50 groups with at least one radio galaxy detected and 23 from the remaining 29 groups. \\ $^b$: 664 from the 50 groups with at least one radio galaxy detected and 213 from the remaining 29 groups.} \label{tab:properties}
\end{table}

The temperature of the groups was determined through the $T_X$ - L$_X$ scaling relation \citep{Giles_2016, Kettula_2015}. A small subsample also had a direct measurement of the temperature available \citep{Kettula_2013}. We find that these measurements are consistent with the scaling relation above.

\begin{figure}
    \includegraphics[height=30em, width=29em, valign=t]{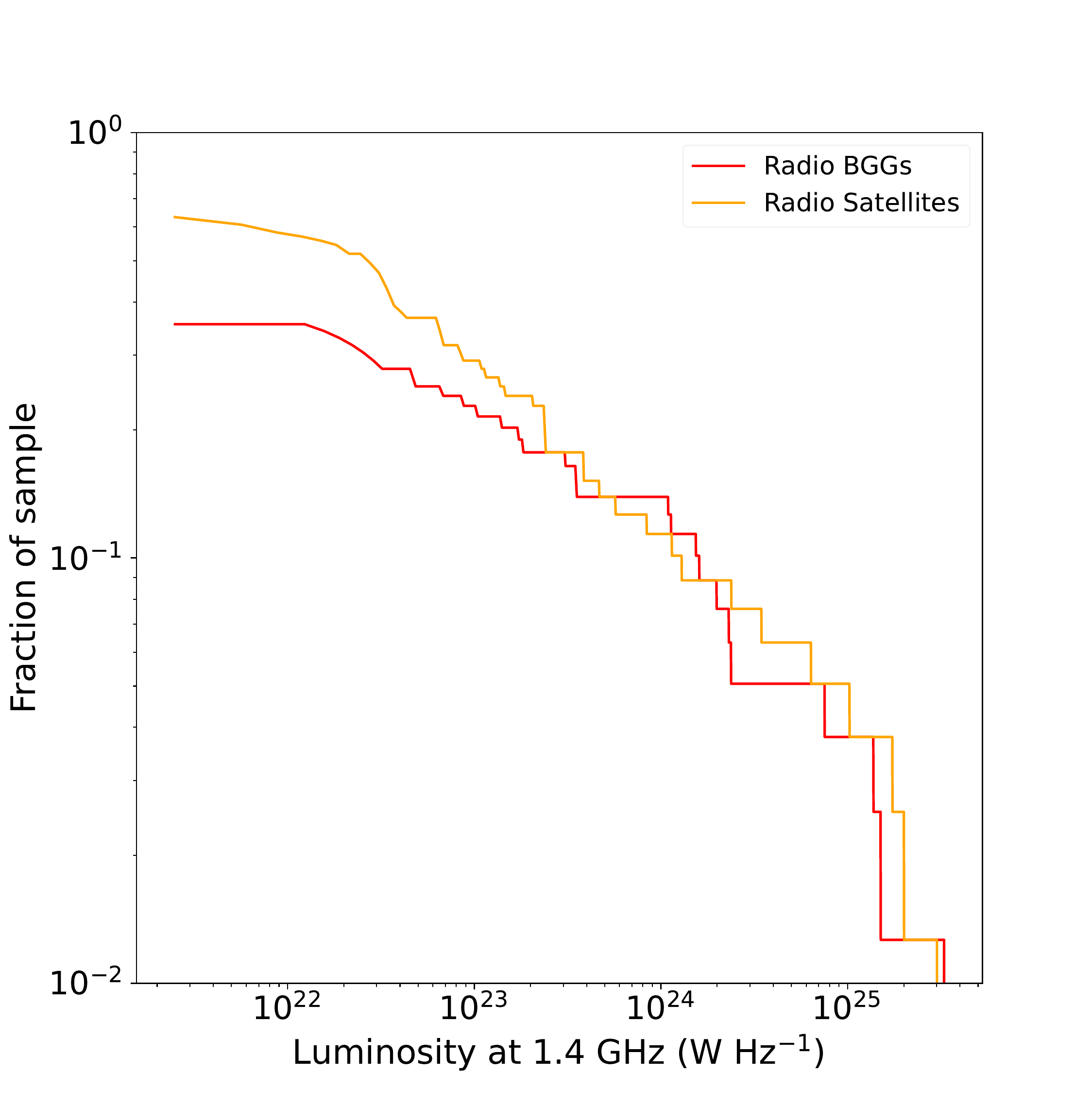}
	\caption{Cumulative 1.4 GHz luminosity distribution for all 79 galaxies with radio emission, divided into BGGs (red) and satellites (orange). The \textit{y-axis} reflects the relative fractions of BGGs (28 out of 79 galaxies) and satellites (51 out of 79 galaxies) that compose the radio galaxy sample.}
	\label{fig:lumfunc}
\end{figure}

The final catalog presents a set of multi-wavelength observables for each group (X-ray luminosity, temperature), for the member galaxies (spectroscopic redshift, velocity dispersion, proper velocity, stellar mass) and for the corresponding radio source, when present (1.4 GHz power, Largest Linear Size).
A selection bias could be introduced by those galaxies - and therefore the host groups - that do not show radio emission according to our criteria. The reason for this could be the lack of an AGN, or limitations set by the sensitivity of VLA-COSMOS. Among all observable galaxies, \citet{Padovani_2017} claim 
that around $\sim$ 1\% host an AGN. Nevertheless, this value should increase in overdense environments such as galaxy groups. \citet{Sabater_2019} report a 100\% detection rate for galaxies with M$> 10^{11}$ M$_\odot$, but this fraction has been observed to strongly vary depending on the host galaxy stellar mass \citep{Kauffmann_2003}. Among our sample, $\sim$ 8\% host an AGN. The radio luminosity distribution for all galaxies with detected radio emission is shown in Fig.~\ref{fig:lumfunc}.

The luminosity distributions span the range $L_R \sim 2 \times 10^{21}$ W Hz$^{-1}$ - $L_R \sim 3 \times 10^{25}$ W Hz$^{-1}$. The end of the range is lower than the BCG luminosity distribution of \citet{Hogan_2015} that reaches $L_R \sim 10^{27}$ W Hz$^{-1}$. The reason for this is that we are only considering galaxy groups, where the radio power of the central galaxy is generally lower than for clusters. This was also found in \cite{Pasini_2020} by investigating the same parent sample used for this work, composed by 247 COSMOS galaxy groups. In \cite{Pasini_2020}, the luminosity reaches $\sim$ 10$^{27}$ W Hz$^{-1}$ in a small ($<5$ out of 247 groups) number of outliers that show very extended AGN emission. The reason why these sources are excluded from the present work is that no spectroscopic data is available for the optical host. Our analysis will therefore be limited to $L_R< 10^{26}$ W Hz$^{-1}$.

At the highest redshift of the sample ($z \sim$ 0.98), the VLA-COSMOS sensitivity at 3$\sigma$ corresponds to $L_{\text{1.4 GHz}}\sim 2 \times 10^{23}$ W Hz$^{-1}$. With the mean redshift setting around $z \sim$ 0.4, we should be able to pick most of the radio sources brighter than 10$^{22}$ W Hz$^{-1}$ at 1.4 GHz. This yields a sample that is representative of the radio luminosity function usually observed for radio galaxies \citep[e.g.][]{Hogan_2015}. Therefore, undetected radio emission should not affect our analysis, which focuses on the comparison between BGGs with detected radio emission (hereafter radio BGGs), satellites with detected radio emission (hereafter radio satellites), BGGs with no radio emission and satellites with no radio emission.

\section{Results and discussion} 

\subsection{Phase-space analysis}
\label{sec:phsp}

\begin{figure*}
    \includegraphics[height=29em, width=55em]{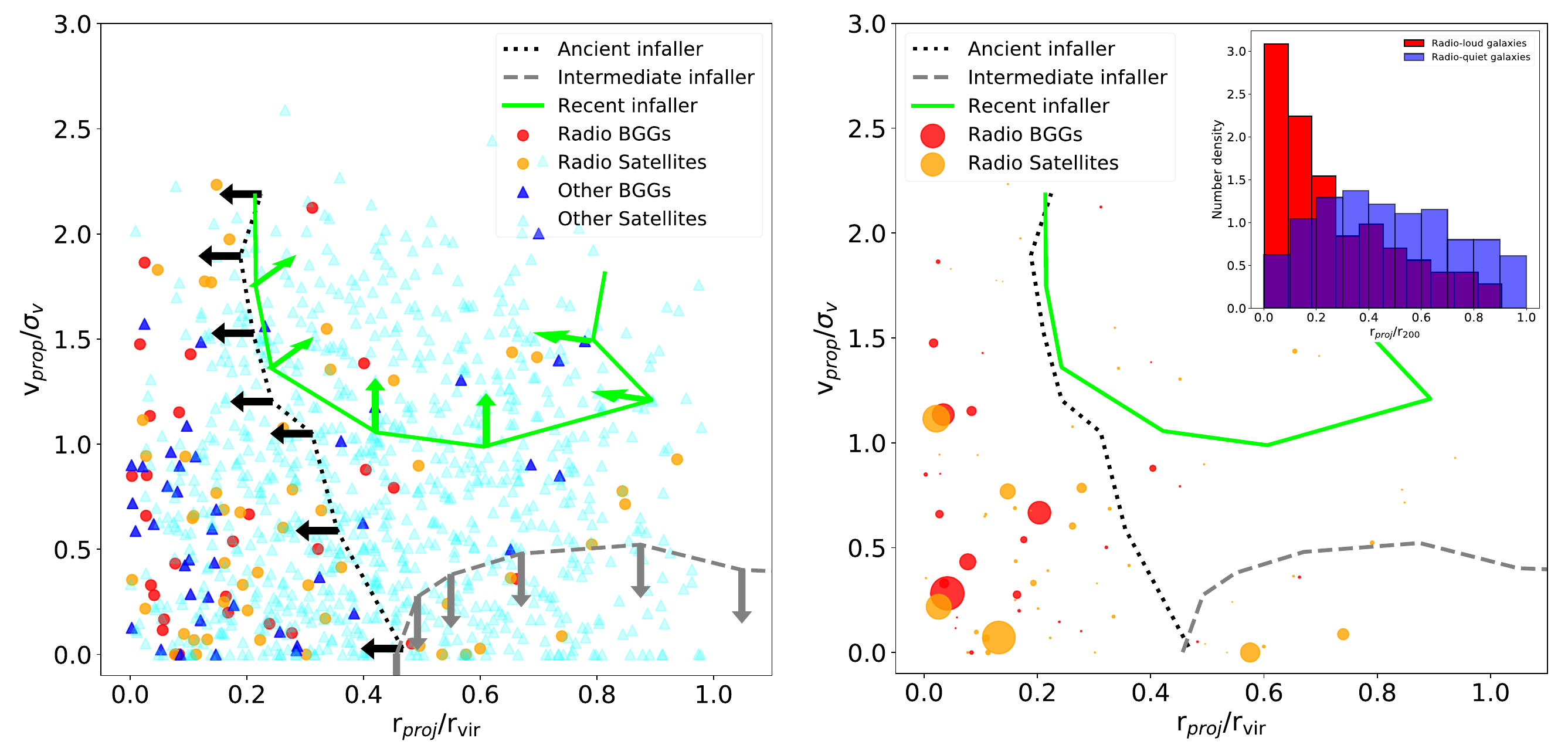}
	\caption{\textit{Left}: Phase-space diagram for radio BGGs (red), non-radio BGGs (blue), radio satellites (orange) and non-radio satellites (cyan). The $x$-axis represents the ratio between the distance from the group centre and $R_{\rm vir}$, while the $y$-axis is the ratio between the line-of-sight velocity and the (one-dimensional) velocity dispersion. The different regions in the diagram indicate ancient infallers (left of black dotted line), intermediate infallers (below the grey dashed line) and recent infallers (above the green line). \textit{Right}: Phase-space diagram restricted to radio BGGs and radio satellites only, with the points sized for the power of the corresponding radio galaxy. The top-right histogram shows the offset distribution for galaxies with (red) and without (blue) radio emission.} 
	\label{fig:phasespace}
\end{figure*}

We performed a phase-space analysis by comparing the cluster/group-centric velocity with the cluster/group-centric offset of the hosted galaxies. This diagram conveys information about the assembly and accretion history of these objects. In the left panel of Fig.~\ref{fig:phasespace} we show the phase-space diagram for radio BGGs, radio satellites, non-radio BGGs and non-radio satellites. 
Following \citet{Rhee_2017} and \citet{Gozaliasl_2020}, the position of each object in this diagram is an indicator of the infall time ($t_{\text{inf}}$) of the galaxy, with ancient infallers (6.45 Gyr $<$ $t_{\text{inf}}$ $<$ 13.7 Gyr) found to the left of the black dotted line in Fig.~\ref{fig:phasespace}, while intermediate infallers (3.63 Gyr $<$ $t_{\text{inf}}$ $<$ 6.45 Gyr) cover the whole offset range below the grey dashed curve. Galaxies above the green line are classified as recent infallers, while the remaining ones cannot be attributed to any of these classes. This does not affect the following analysis since our purpose is to distinguish ancient infallers from all the other objects. Out of 28 radio BGGs, only 5 ($\sim$ 18\%) are not classified as ancient infallers, which constitute $\sim$82\% of the sample. On the other hand, the sample of 42 non-radio BGGs is composed of 33 ancient infallers ($\sim$78\%), consistently with \citet{Gozaliasl_2019}, who also found that BGGs are mostly ancient infallers. The sample of radio satellites show $\sim$65\% ancient infallers. Finally, only $\sim$41\% of non-radio satellites present this classification.


The right panel of Fig.~\ref{fig:phasespace} shows objects with radio emission only (BGGs and satellites), with the size of the symbols proportional to the power of the radio source. The top-right histogram shows the offset distribution for galaxies with and without radio emission. The comparison between the radio and non-radio samples in the histogram clearly indicates that most of the galaxies with radio emission are ancient infallers (56 out of 79, $\sim$ 71\%), strongly peaking at low offsets, while the distribution of the distances from the group centre for galaxies with no radio emission is more uniform across $R_{\rm vir}$. The phase-space analysis applied to radio objects-only also suggests that powerful radio galaxies ($L_R> 10^{23}$ W Hz$^{-1}$) are always located close to the group centre ($< 0.2\,R_{\rm vir}$). This is expected since central galaxies switch the SMBH on much easily. The gas cooled out of the Intra-Group Medium (IGrM) can feed the AGN if the galaxy lies close to the group density peak, where the cooling is more efficient. Nevertheless, galaxies located in the outskirts or outside the cooling radius of the group can still show radio emission. However, they might have to rely on more episodic triggers, such as mergers or interactions with other objects. Only in a few cases, their radio power is able to become comparable to those of central galaxies. This happens especially because the low density in the outskirts of galaxy groups sometimes allow them to grow rapidly in size (see also \citealt{Pasini_2020}).

A further consequence of this is that radio BGGs have a higher chance than non-radio BGGs to lie close to the group centre. This is particularly true for the most powerful ones ($L_R> 10^{23}$ W Hz$^{-1}$), that in our samples always lie within 0.2\,$R_{\rm vir}$. Therefore, the detection of a powerful radio source in a group can help identify the group centre. Finally, it is worth noting that no difference is visible in terms of velocity ratio between powerful BGGs and those with $L_R< 10^{23}$ W Hz$^{-1}$.


\subsection{Comparison with cosmological simulations}
\label{sec:hzsec}

\begin{figure*}
    \hspace{-0.1cm}
	\includegraphics[height=31.5em, width=29.5em, valign=t]{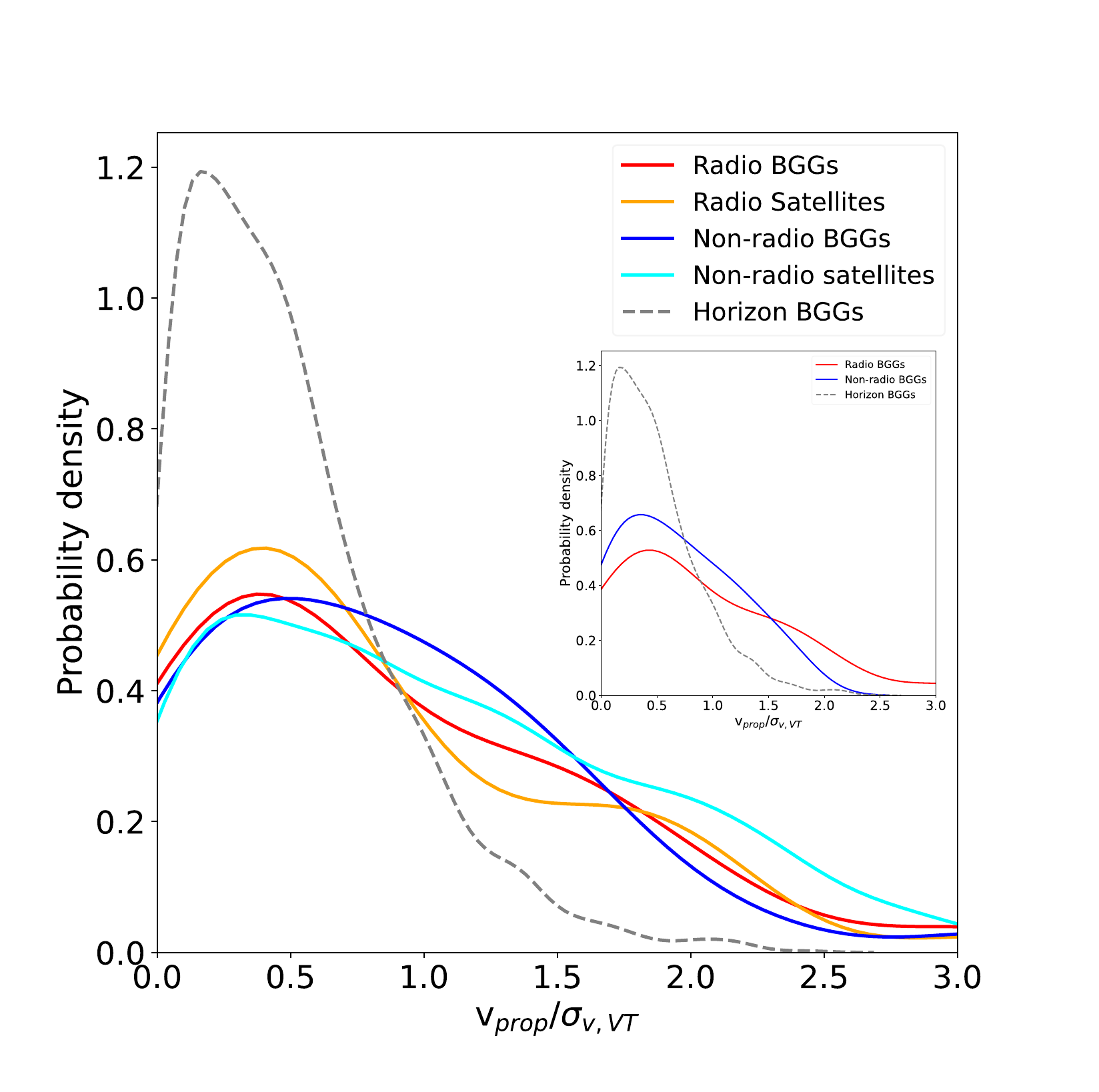}
	\includegraphics[height=31em, width=29.5em, valign=t]{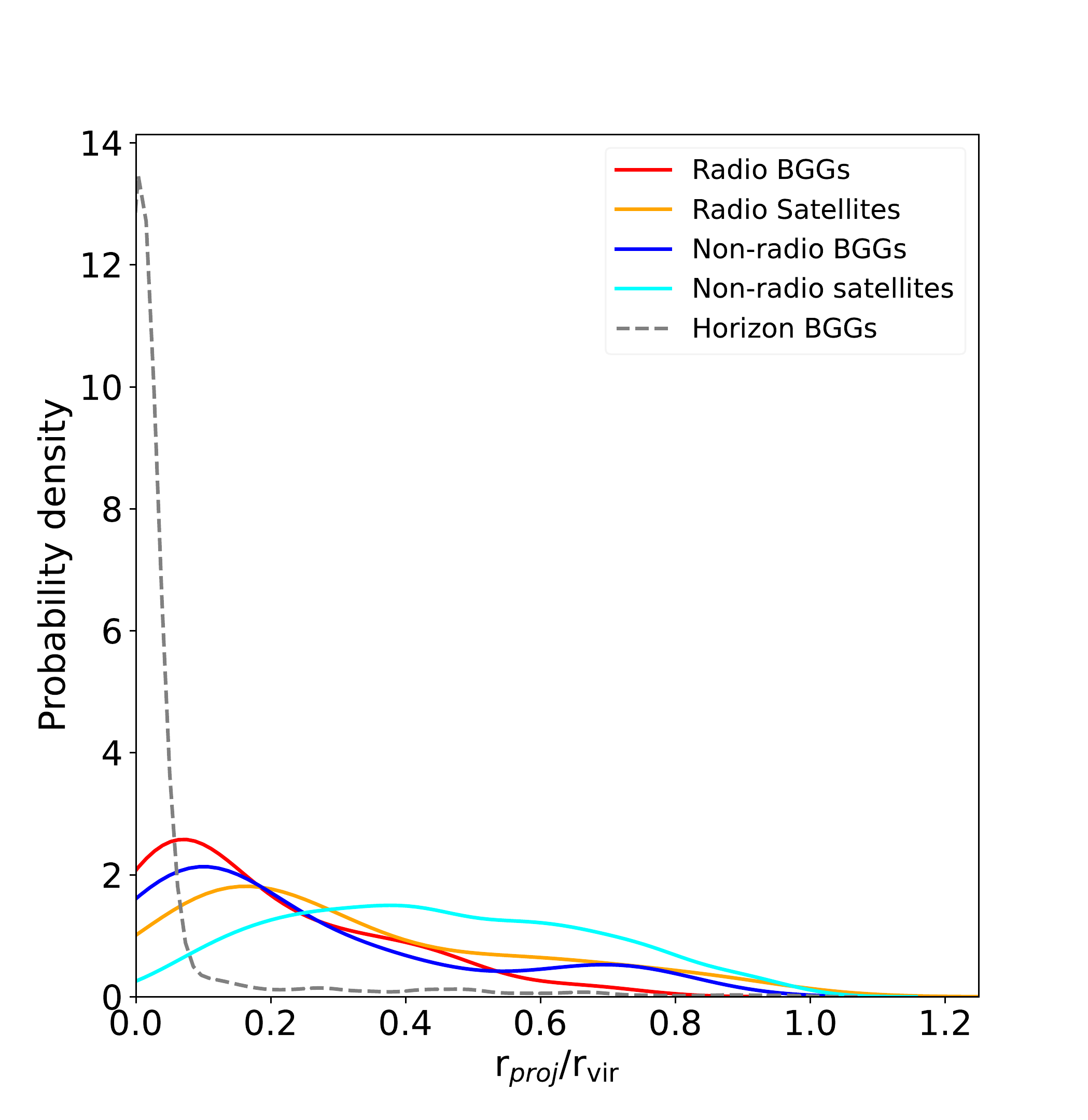}
	\caption{\textit{Left}: Probability density (obtained through a gaussian kernel density estimate) vs. velocity ratio for radio and non-radio BGGs (red and blue), for radio and non-radio satellites (orange and cyan) and for HZ-simulated BGGs (dashed gray).The subplot shows the same distribution restricted to only BGGs with offset $<$ 0.3$R_{\rm vir}$. \textit{Right}: Probability density vs. offset from the group centre for radio and non-radio BGGs (red and blue), for radio and non-radio satellites (orange and cyan) and for HZ-simulated BGGs (dashed gray).} 
	\label{fig:plotdens}
\end{figure*}

Here, we compare our samples with the theoretical predictions from the HORIZON-AGN (HZ) simulation\footnote{https://www.horizon-simulation.org} \citep{Dubois_2014}. The HORIZON-AGN simulation is a cosmological hydrodynamical simulation of 100 Mpc/$h$ comoving box containing $1024^3$ Dark Matter particles. The simulation is performed with the adaptive-mesh refinement code RAMSES \citep{Teyssier_2002} including gas dynamics, gas cooling and heating, and sub-grid models for star formation, stellar and AGN feedback. The AdaptaHOP halo finder \citep{Aubert_2004} was run on both the stellar and DM particle distributions to identify galaxies and halos (see \citealt{Laigle_2019} and \citealt{Gozaliasl_2019} for further details). Each galaxy is then associated with its closest main halo. To match the observational definition, the BGG is identified as the most massive galaxy within the virial radius of the main halo. 

The left panel of Fig.~\ref{fig:plotdens} shows the probability density distribution of the velocity ratio for radio and non-radio BGGs and for radio and non-radio satellites. Here, instead of the gapper velocity dispersion $\sigma_{\mathrm v}$ exploited in Fig. \ref{fig:phasespace}, we use the velocity dispersion estimated from X-ray emission $\sigma_{\mathrm v, VT}$ (see below). The gray dashed curve represents the distribution for simulated BGGs obtained by HZ-AGN, whose mass and redshift evolution were already studied and compared to COSMOS BGGs and satellites in \citet{Gozaliasl_2020}. Here, our purpose is to understand whether the dynamical properties of our four samples differ significantly from each other, and how they compare to simulated galaxies.

The distributions of radio and non-radio BGGs peak at $v_{\text{prop}}/\sigma_{\rm v, VT} \sim 0.4$ and 0.5, respectively. For radio BGGs, the mean velocity ratio is $\sim$ 0.84 and the median velocity ratio is $\sim$ 0.63. For non-radio BGGs the mean velocity ratio is $\sim$ 0.85 and the median is $\sim$ 0.78. Radio satellites have the peak velocity ratio around $\sim$ 0.45, with a mean of $\sim$ 0.78 and a median of $\sim$ 0.59. The distribution for non-radio satellites is broader, setting the peak at $\sim$ 0.3 but becoming the dominant sources at $v_{\text{prop}}/\sigma_{\rm v, VT} \geq 1.6$, with a mean of $\sim$ 1.04 and a median of $\sim$ 0.89. Finally, BGGs in the HZ-AGN simulation are strongly peaked around $v_{\text{prop}}/\sigma_{\rm v, VT}$ $\sim$ 0.2, with a mean of $\sim$ 0.49 and a median of $\sim$ 0.41. 

Given that the simulation represents the dynamics of the central galaxy, in the subplot of Fig. \ref{fig:plotdens} we select only BGGs within 0.3$R_{\rm vir}$ from the X-ray center, to see if this changes the observed displacement of the curve with respect to HZ BGGs. We see no difference in the distribution for radio BGGs, while the curve for non-radio BGGs becomes tighter. Nevertheless, the distribution for HZ BGGs remains much more strongly peaked at low velocity with respect to real BGGs, with a steeper decrease after the peak.

The probability distribution of offsets from the group centre, shown in the right panel of Fig.~\ref{fig:plotdens}, confirms that BGGs (radio, non-radio and simulated) are highly concentrated within 0.2 $R_{\rm vir}$. At higher radii the curves of radio and non-radio BGGs decrease in a similar fashion, while the probability density of simulated BGGs already steepens at $\sim$ 0.1 $R_{\rm vir}$. Nevertheless, in simulations BGGs tend to be more massive than their observational counterparts \citep[e.g.,][]{Bahe_2017, Henden_2019, Bassini_2020}. For this reason, they are closer to the centre and exhibit less spread in velocity. On the other hand, satellites exhibit a broader distribution, with radio satellites prevailing at offsets $\leq 0.3$, while at outer radii non-radio satellites become dominant. 

All BGGs were observed as a part of zCOSMOS survey \citep{Lilly_2007}, with a redshift error of $\sim$55 km/s. Missing objects were covered by FORS2 program \citep{George_2011}, with a similar redshift precision, and at $z>0.7$ by GEEC2 \citep{Balogh_2011}, with a redshift precision of 80 km/s.
Based on the work of \citet{Saro_2013},  the uncertainty of velocity dispersion measurement is high with typical number of spectroscopic members of COSMOS X-ray galaxy groups. Better constraints on the velocity dispersion are obtained using scaling relations of $L_X-M_{200}$ \citep{Leauthaud_2010} and $M_{200}-\sigma_{\rm v, VT}$ \citep{Carlberg_1997}. The log-normal scatter $L_X-\sigma_{\rm v, VT}$ relation is measured to be 0.13 by \citet{Kirkpatrick_2021}. Using the velocity dispersion from scaling relations, the disagreement with simulations consists in a wider tail above $0.7\sigma$ extending to $2.5\sigma$. The spread due to uncertainty in the mean redshift is typically 0.3$\sigma_{\rm v, VT}$ and always better than 0.45$\sigma_{\rm v, VT}$ and cannot explain the large tail.


Positional displacement of BGGs from the center of the halo is constrained to be within 0.1 $R_{\rm vir}$ in simulations, while our data shows much broader range of offsets between BGG and X-ray peak. \citet{George_2012} found that BGGs in the vicinity of X-ray centers (within 0.25 $R_{\rm vir}$) are good tracers of projected mass centers, with an offset less than 0.1 $R_{\rm vir}$, but BGGs with strong offsets from X-ray center do not trace the center of mass and the corresponding mass profiles suggest merging. Thus, the broad distribution of offsets within 0.3 $R_{\rm vir}$ is due to displacements of X-ray peak, while larger offsets are merger driven. For the offset peak of HZ galaxies to match that of our data, it would require a systematic shift of $\sim$0.1 $R_{\rm vir}$ (10$\arcsec$ - 6$'$).

To quantify our results, we performed a Kolmogorov-Smirnov (KS) test to compare the radio and non-radio BGG distributions with simulated BGGs. Our null-hypothesis is that the samples are drawn from the same parent distribution. The KS-test on the phase-space distributions of radio BGGs and simulated BGGs gives a null-hypothesis probability of $p = 1.6 \times 10^{-13}$, while the comparison between non-radio BGGs and simulated BGGs results in $p = 3.1 \times 10^{-17}$. This suggests that the simulation is not able to reproduce our samples, indicating that it may need additional physics to reproduce the true population of BGGs.

\subsection{Properties of the samples}

\begin{figure*}
    \hspace{-0.1cm}
	\includegraphics[height=29.5em, width=29.5em]{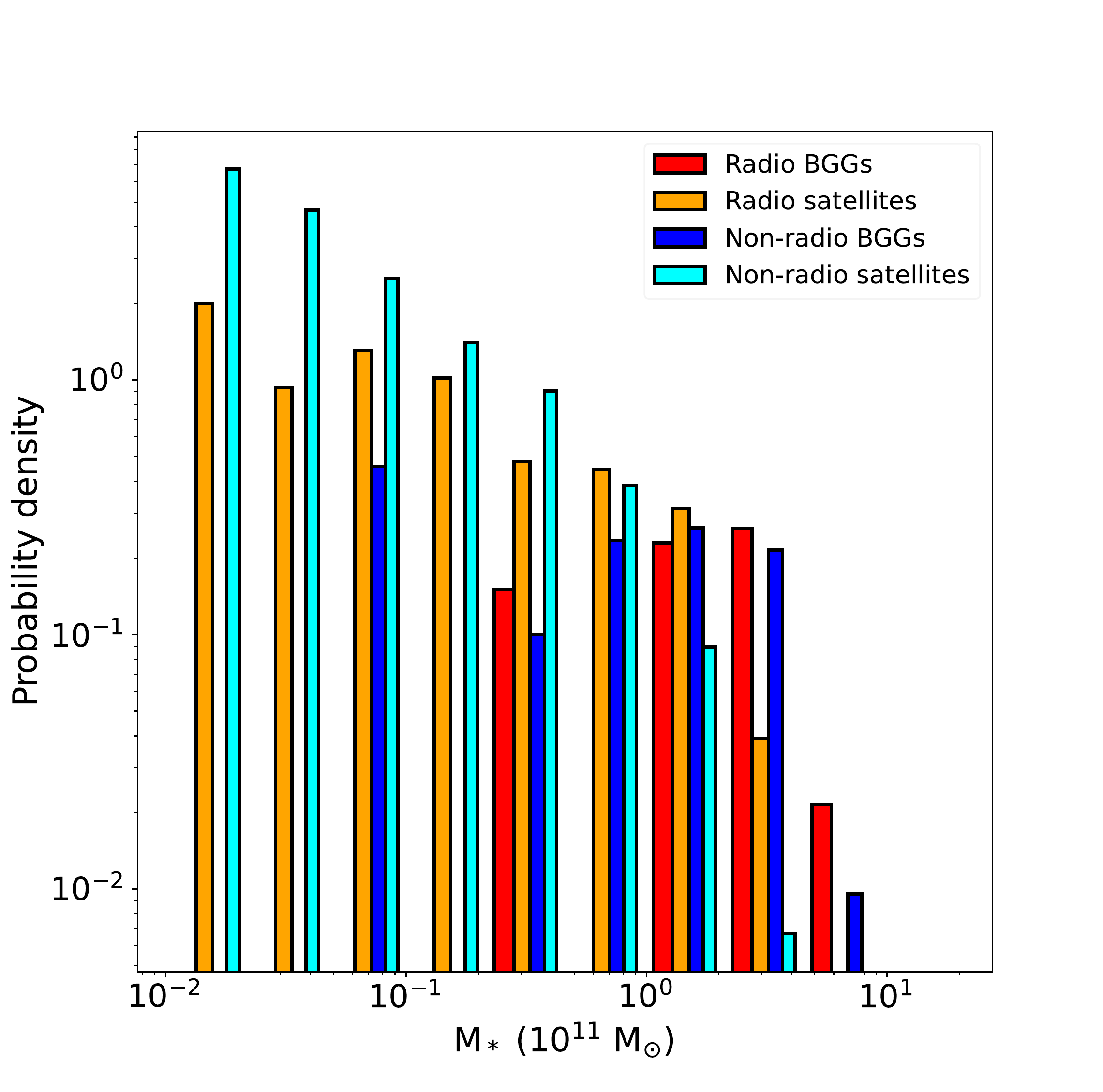}
	\includegraphics[height=29.5em, width=29.5em]{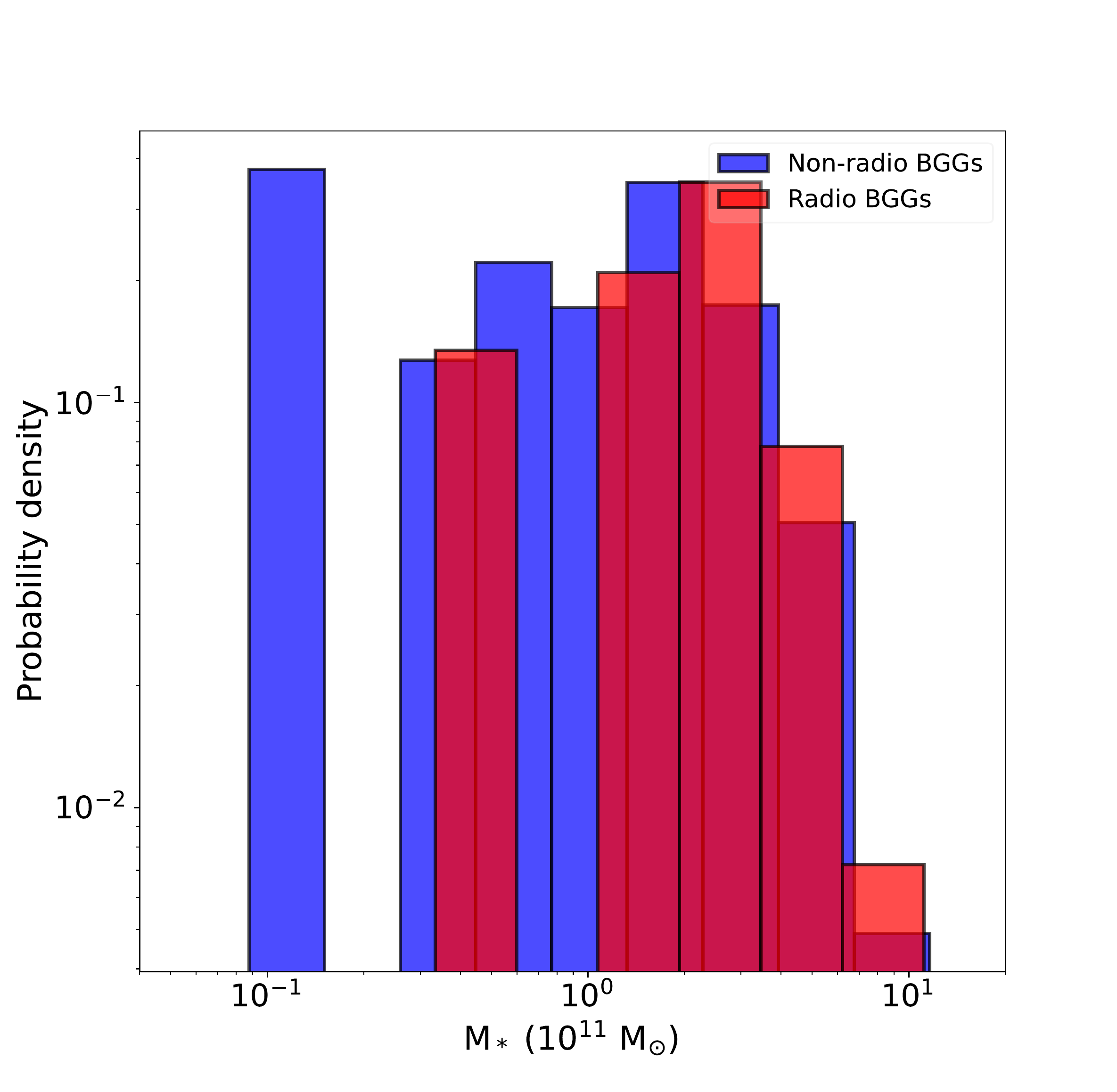}
	\caption{\textit{Left}: Stellar mass distribution for radio (red) and non-radio (blue) BGGs, and for radio (orange) and non-radio (cyan) satellites. \textit{Right}: Stellar mass distribution restricted to radio (red) and non-radio (blue) BGGs.} 
	\label{fig:massdistr}
\end{figure*}

The left panel of Fig.~\ref{fig:massdistr} shows the stellar mass distribution for the four samples: radio BGGs, radio satellites, non-radio BGGs and non-radio satellites. Non-radio satellites dominate the low-mass regime, from 10$^9$ M$_\odot$ to $5 \times 10^{10}$ M$_\odot$. Low-mass radio BGGs start to appear around $3 \times 10^{10}$ M$_\odot$, while non-radio BGGs go down to 10$^{10}$ M$_\odot$. Radio BGGs become dominant for stellar masses $\geq 3 \times 10^{11}$ M$_\odot$. In order to compare the mass distributions of the BGG samples,  we show the histograms of radio and non-radio BGGs in the right panel of Fig.~\ref{fig:massdistr}. Out of 28 radio BGGs, 27 have $M_* > 10^{11}$ M$_\odot$ ($\sim$96\%), while the same fraction for non-radio BGGs is $\sim$83\% (35 out of 42). The small size of the samples does not lead to statistically significant results, but radio BGGs trend towards higher masses, while it is less likely for a BGG with $M_* < 10^{11}$ M$_\odot$ to host a radio galaxy.

\begin{figure}
    \hspace{-0.1cm}
	\includegraphics[height=29.5em, width=29.5em]{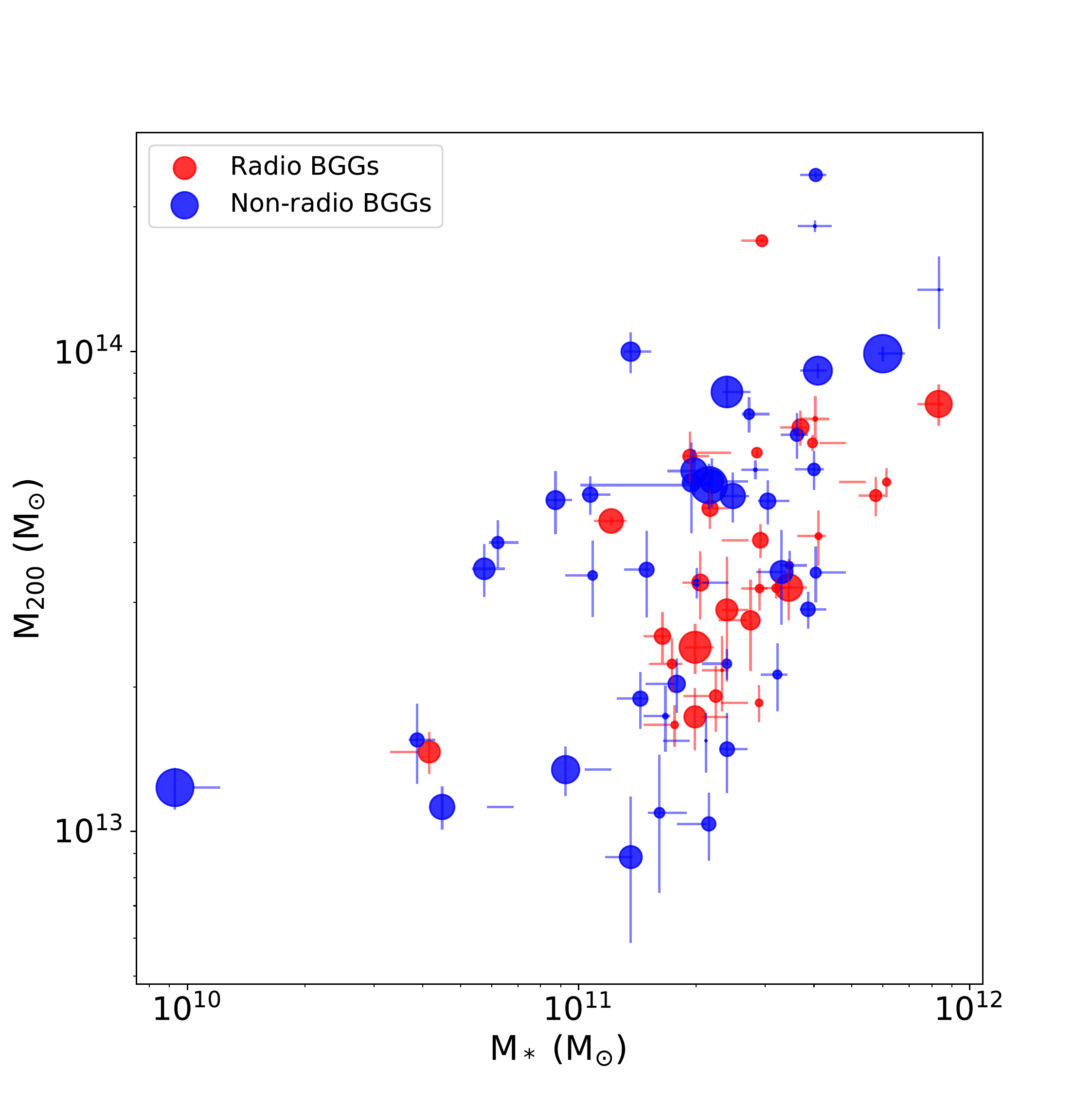}
	\caption{Galaxy groups $M_{200}$ vs. stellar mass for the radio-BGG (red) and non-radio BGG (blue) samples. The size denotes the distance of the BGG from the group centre, with bigger points indicating larger offsets.}
	\label{fig:mstarmgroup}
\end{figure}

This is even more significant when considering that massive BGGs are found in more massive groups and vice versa, as discussed in several papers \citep[e.g.,][]{Stott_2010, Gozaliasl_2016} and shown in Fig.~\ref{fig:mstarmgroup}. This is in agreement with \citet{Gaspari_2019} who observed that more massive SMBHs correlate with larger and hotter X-ray halos. The same is also found in cosmological simulations by, e.g.,  \citet{Bassini_2019, 2021MNRAS.501.2210T}. In combination with the result shown above that radio BGGs are usually more massive than those with no radio emission, this suggests that radio BGGs are more likely to be hosted in high-mass groups. Since radio-BGGs statistically exhibit smaller offsets from the centre than non-radio BGGs (see Sec.~\ref{sec:phsp}), we can argue that this could affect the trigger of the AGN. Indeed, Fig.~\ref{fig:phasespace} shows that a number of non-radio BGGs have large offsets. This is not surprising considering that the group centre as we defined it, i.e., the bottom of the potential well, is where the hot gas density is higher and cooling is faster, especially in more relaxed systems. Stronger cooling implies more condensing mass (\citealt{Gaspari_2019}) which feeds the supermassive black hole (SMBH) (e.g., see the GR-rMHD simulations by \citealt{Sadowski_2017}). Nevertheless, there are also high-mass non-radio BGGs which do not host a radio source, at least with our sensitivity limit. This could likely be explained by the flickering duty cycle involved in the AGN feeding and feedback self-regulation (Sec.~\ref{sec:discussion}) .



\begin{figure}
	\includegraphics[height=30em, width=29.5em]{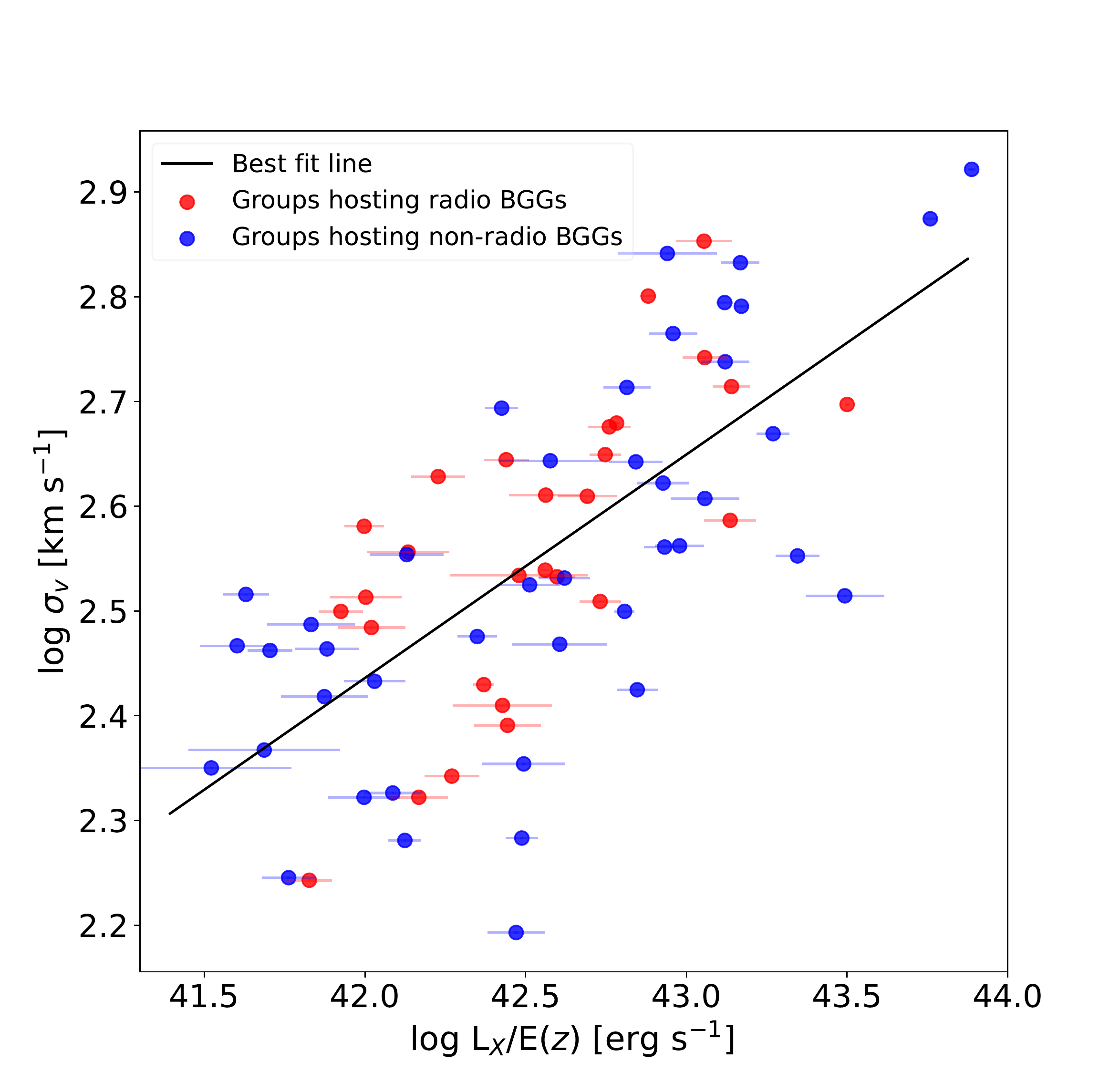}
	\caption{Observed velocity dispersion of the host group vs. X-ray luminosity for radio (red) and non-radio (blue) BGGs. The black line represents the scaling relation obtained by computing the L$_X$ - M and the M - $\sigma_{v, VT}$ correlations presented in \citet{Leauthaud_2010} and \citet{Mamon_2013}, respectively.} 
	\label{fig:scalrel}
\end{figure}

In order to determine whether one of our samples shows any divergence from standard scaling relations, we investigated the correlation between X-ray luminosity and observed velocity dispersion ($\sigma_{\rm v}$) for groups hosting radio and non-radio BGGs, plotted in Fig.~\ref{fig:scalrel} \citep[e.g.,][]{Wu_1999, Mahdavi_2001, Zhang_2011, Gozaliasl_2020}. None of the samples seem to show any deviation and both follow the same correlation. This is confirmed by the KS test ($p$ = 0.06), which suggests that the two distributions are similar and that no discernible difference exists in the $\sigma_{\rm v}$-L$_X$ correlation between the groups hosting radio and non-radio BGGs.
  
We also looked for evidence of recent interactions with other galaxy groups by inspecting the \textit{Chandra} and XMM-Newton observations used to build the original galaxy groups catalog \citep{Gozaliasl_2019}. We find that only for one group it is possible to detect hints of mergers (LSS 17, see \citealt{Smolcic_2007}). The images are too shallow to reveal anything for the other objects. We will return to this issue in Sec.~\ref{sec:discussion}, but deeper X-ray observations will be needed for this purpose.

It is well-known that the magnitude difference between the first- and second-rank galaxies in a group/cluster is helpful to trace their merger history and evolution \citep{Ponman_1994, Gozaliasl_2014, Gozaliasl_2019}. Simulations have shown that mergers in galaxy groups lead to runaway growth of the BGG \citep[e.g.,][]{Cavaliere_1986, Mamon_1992}, at the expense of the second brightest galaxy. Therefore, the magnitude gap between the BGG and the second-rank galaxy should increase in time, finally leading to a situation in which the central, elliptical BGG, lying in an X-ray luminous halo, is surrounded by faint satellites. These  groups are known as \textit{fossil groups} \citep{Jones_2003}. Therefore, one expects larger gaps in more relaxed groups, where the offset between the BGG and the halo centre is lower. The magnitude gap - offset relation for our parent sample of galaxy groups has already been studied in \citet{Gozaliasl_2019}. Here, we wish to understand how this gap relates to BGGs with and without radio emission.

\begin{figure}
	\includegraphics[height=29.5em, width=29.5em]{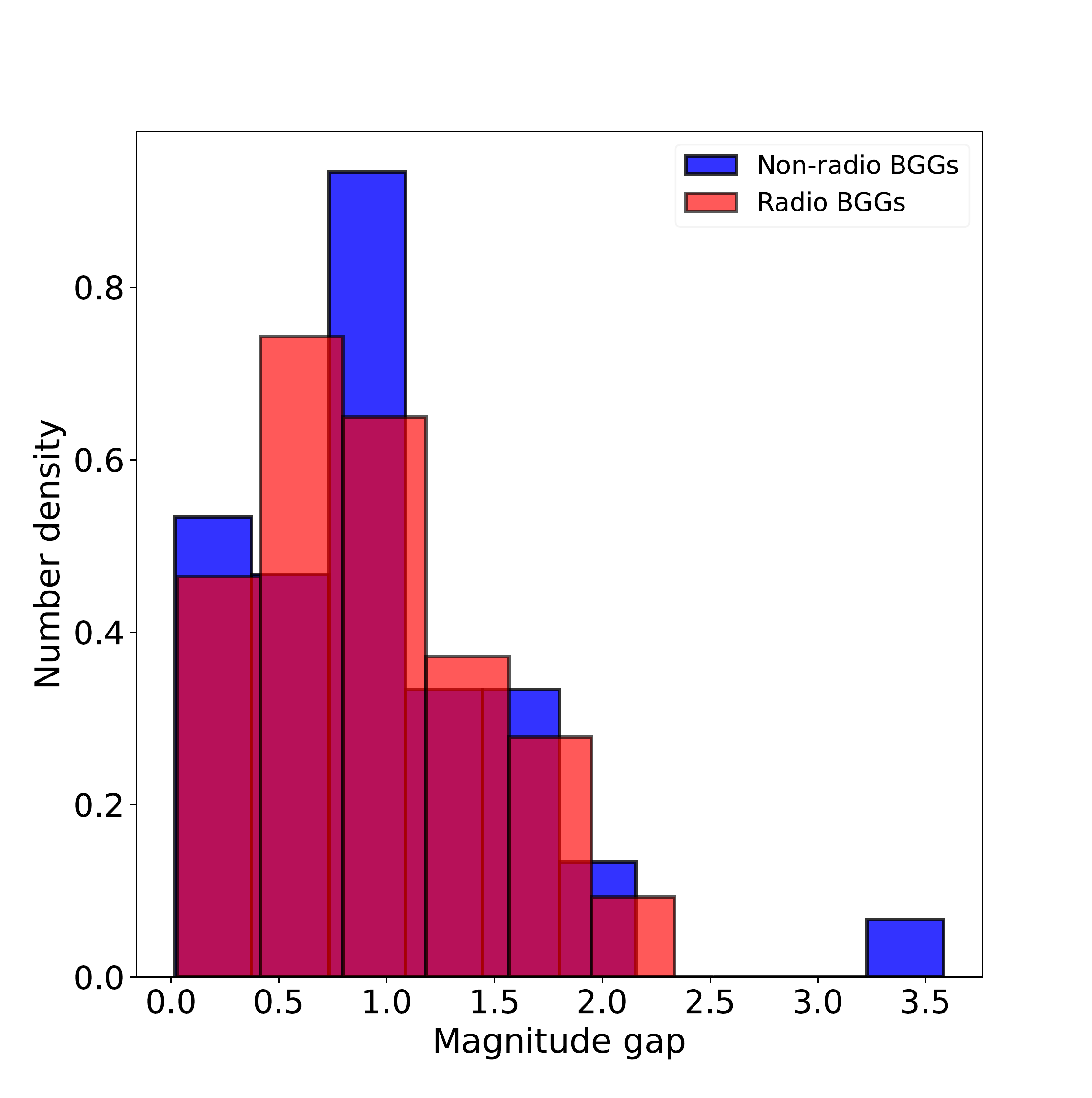}
	\caption{Magnitude gaps (R-band) of radio (red) and non-radio (blue) BGGs from the corresponding second-rank galaxy.} 
	\label{fig:maggap}
\end{figure}

In Fig.~\ref{fig:maggap}, we show the R-band magnitude gap distribution for our  brightest groups galaxies, classified into radio and non-radio BGGs. The gap was measured within 0.5 $R_{200}$, following \citep{Jones_2003}. The second-rank galaxy of each group was picked, regardless of it having spectroscopic or photometric identification, using the full membership catalog studied in \citet{Gozaliasl_2020}. Both distributions peak around $\sim 1$, with a clear concentration of objects before $\sim$1.5. No clear difference in the trend is detected between radio and non-radio BGGs, as also confirmed by the KS test ($p =  0.13$). Nevertheless, our catalogs only have a few tens of objects, and analyses on larger samples could help to address how the magnitude gap relates to radio and non-radio BGGs.

\subsection{Scaling relations}
\label{scalings}

\begin{figure*}
	\includegraphics[height=31em, width=29em]{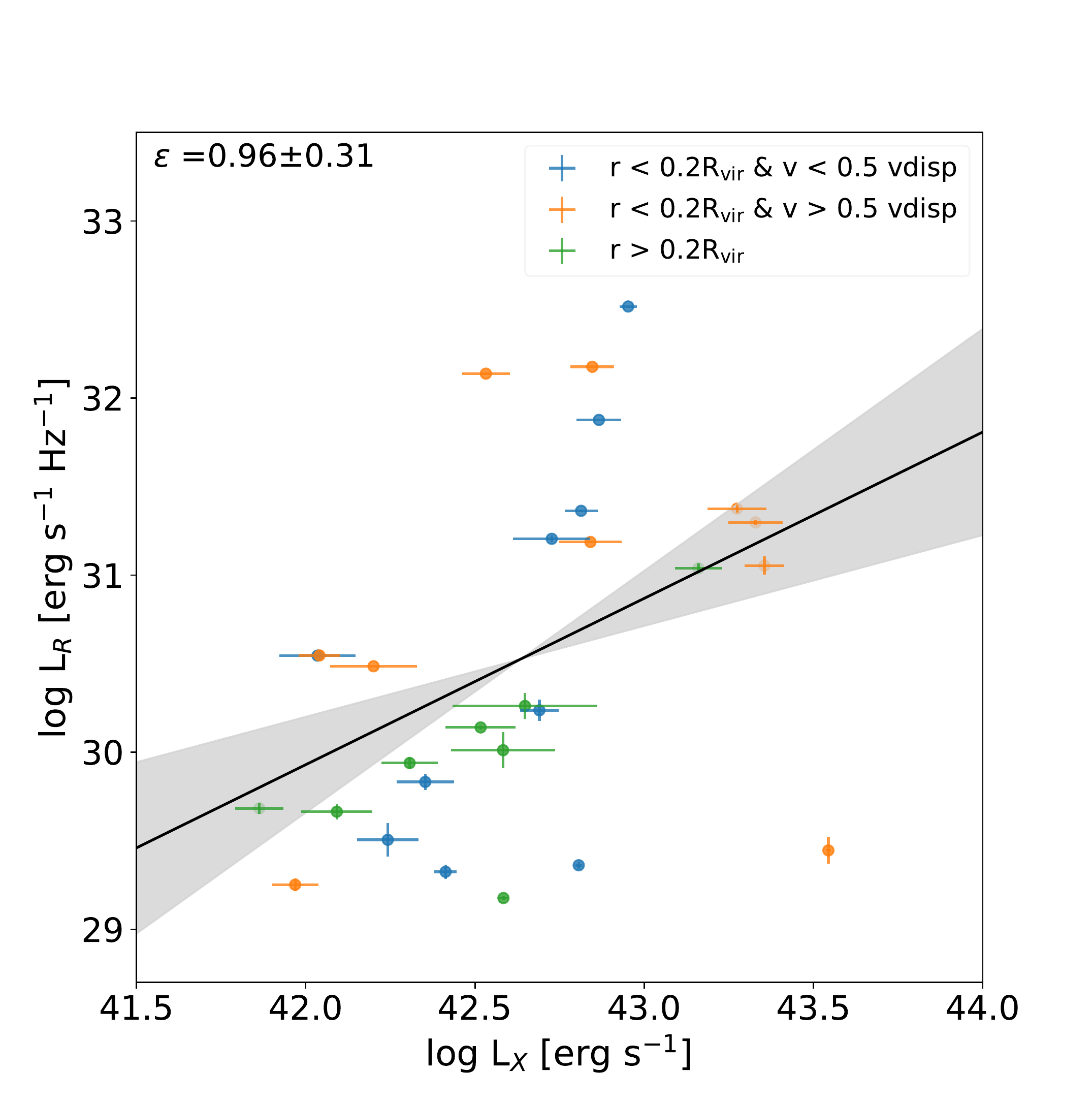}
	\includegraphics[height=31em, width=29em]{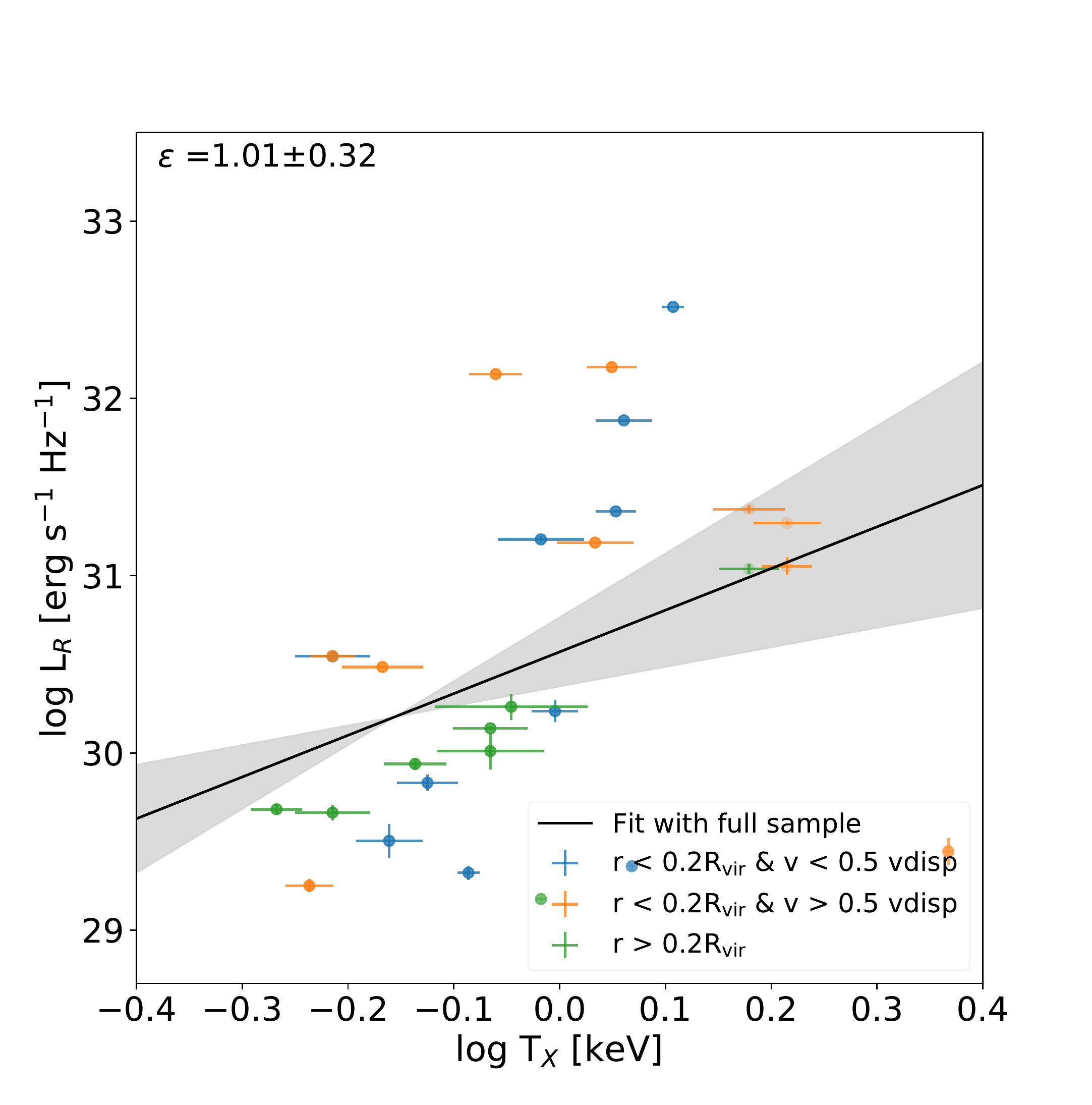}
	\caption{\textit{Left:} 1.4 GHz AGN radio power vs. X-ray luminosity of the host group for the radio BGG sample. The sample was divided into relaxed objects (cyan, offset $< 0.2\,R_{\rm vir}$ and $v< 0.5 v_{\text{disp}}$), high-velocity objects (yellow, offset $< 0.2\,R_{\rm vir}$ and $v > 0.5 v_{\text{disp}}$) and offset objects (green, offset $> 0.2\,R_{\rm vir}$). The black line represents the best-fit relation obtained through a Bayesian statistical analysis (the intrinsic scatter is labeled in the top-left corner). 
	\textit{Right:} 1.4 GHz AGN radio power vs.~temperature of the host group. The classification is the same as in the left panel. The grey line shows 1$\sigma$ errors on the best fit.}
	\label{fig:radiox}
\end{figure*}

\citet{Pasini_2020} explored a correlation between the 1.4 GHz radio power of the central AGN and the X-ray luminosity of the host galaxy group. Since the sample of radio BGGs studied in this work is derived from the same parent catalog of X-ray groups \citep{Gozaliasl_2019}, we expect to find a similar correlation, which is shown in the left panel of Fig.~\ref{fig:radiox}.

The sample was divided into relaxed objects (offset $< 0.2\,R_{\rm vir}$ and $v< 0.5 \,v_{\text{disp}}$), high-velocity objects (offset $< 0.2\,R_{\rm vir}$ and $v > 0.5 \,v_{\text{disp}}$) and offset objects (offset $> 0.2\,R_{\rm vir}$). We find no significant differences between the three subsamples. We then applied Bayesian inference to extract the best-fit relation, by using {\ttfamily linmix\footnote{https://github.com/jmeyers314/linmix.}} (cf.~Sec.~2.2 in \citealt{Gaspari_2019} for a discussion of its features). A linear fit in log-log scale was performed in the form:

\begin{equation}
    Y = \alpha + \beta X + \epsilon,
\end{equation}
with $\alpha$ and $\beta$ representing the intercept and the slope, respectively, while $\epsilon$ is the intrinsic scatter of the relation. For the AGN power versus X-ray luminosity correlation, we find $\alpha = -9.53 \pm 18.19$, $\beta = 0.94 \pm 0.43$ and $\epsilon = 0.96 \pm 0.31$. As expected, this estimate is consistent with the slope of $L_R \propto (1.07 \pm 0.12) L_X$ presented in \citet{Pasini_2020}, obtained through least-squares linear regression. However, we note that the errors are large due to the small sample size. 
It is generally understood that the X-ray emission by clusters and groups is tightly linked to the temperature of the ICM \citep{Lovisari_2020}. Therefore, if a correlation of the hot plasma with the AGN exists, we expect a similar link with the gas temperature, too.
The right panel of Fig.~\ref{fig:radiox} shows the correlation between the 1.4 GHz AGN power and the X-ray temperature of the intragroup medium of the host. Again, no  difference is apparent among the subsamples. We find $\alpha = 30.57 \pm 0.19$ and $\beta = 2.35 \pm 1.25$, with intrinsic scatter $\epsilon = 1.01 \pm 0.32$.

The significant positive and steep correlation with X-ray halo properties can be compared with those by \citet{Gaspari_2019}, who found also key positive X-ray halo correlations between the (direct/dynamical) SMBH masses and the observed $T_{\rm x}$ and $L_{\rm x}$. Specifically, their group-dominated sample show a slope $M_{\rm BH} \propto T_{\rm x}^{2.14\pm0.25}$, which is well consistent with the above mean relation. 
This suggests that, despite the AGN power being an instantaneous measure ($L_{\rm R} \sim P_{\rm BH} \propto \dot M_{\rm BH}$), the mean $L_{\rm R} - T_{\rm x}$ is not drastically altered by the details of the feedback duty cycle, except by introducing a larger intrinsic scatter ($4\times$) due to the chaotic intermittency. Such a variable duty cycle is a feature corroborated by a wide range of numerical and observational studies (e.g., \citealt{McNamara-Nulsen_2007,Gaspari_2011b,Fabian_2012,Prasad_2015,Yang_2016}).
The $L_{\rm R} - L_{\rm x}$ relation appears to be slightly steeper than the $M_{\rm BH} - L_{\rm x}$, although still comparable within the 1-$\sigma$ uncertainty.
Overall, as we will discuss in Sec.~\ref{sec:discussion}, a hotter halo implies a larger gas mass, stronger CCA feeding and stronger AGN feedback power, thus establishing major positive correlations. In this regard, the $L_{\rm R} - T_{\rm x}$ relation can be used as a proxy to describe AGN feedback and feeding rates. 

In passing, it is worth noting that all BGGs found at distances of more than 0.2\,$R_{\rm vir}$ from the X-ray centre (green) lie underneath the mean best-fit line suggesting that, at a given X-ray luminosity (or temperature), their radio luminosity is lower than in the more central BGGs. Again, the centeredness of the system appears to be key to initiate stronger feedback (and related feeding; see next Sec.~\ref{sec:discussion}).

\subsection{AGN feeding and feedback cycle}

\label{sec:discussion}

In the previous sections, we showed that AGN activity is usually detected at the centres of galaxy groups, regardless of the properties of the optical galaxy. The triggering of the (mechanical) AGN activity thus appears to depend on the position of the host halo. 
In CCA, the central position in the group promotes the condensation and inflow of low-momentum gas from both the internal galactic gas and external intragroup medium. As a consequence, in CCA the AGN radio power correlates with the X-ray halo temperature (and luminosity), consistent with the results in Sec.~\ref{scalings}. 
During CCA, inelastic collisions between the condensed clouds and filaments drive a rapid inflow toward the micro scale (a few tens of the Schwarzschild radius; \citealt{Gaspari_2017}). The ultrafast outflows and jets then entrain gas at the meso-scale ($\sim 1$ kpc), with their kinetic energy dissipated and released at the macro-scale (tens kpc) via X-ray bubbles, shocks, and turbulence (e.g., \citealt{Gaspari_2011b,Barai_2016,Yang_2019,Liu_2019,Wittor_2020}).
Thus, the ensuing cooling flow of the central intragroup medium can be quenched rapidly by AGN heating, leading to a new feedback cycle (\citealt{Gaspari_2020} for a review and unification diagram of such processes). 
On the other hand, the AGN in non-centrals/'satellites' can only feed from the diffuse gas inside its host galaxy (`internal weather'), and cannot easily tap into the reservoir of the intragroup medium due to their large infall/relative velocity. This is supported by our results, as powerful radio galaxies  mostly lie at the group centre. 

An alternative accretion mode for the SMBH is hot accretion. This mode can take various forms, from pure Bondi accretion -- usually based on idealized assumptions such as the presence of a spherically symmetric, steady, adiabatic and gaseous atmosphere \citep{Bondi_1952} -- to Advection-Dominated Accretion Flow (ADAF, \citealt{Narayan_1995}). 
Unlike in CCA, in hot-mode accretion, the gravitational pull of the SMBH is strongly counterbalanced by the thermal pressure of the hot X-ray halo, which needs to be overcome to allow the inner SMBH feeding.
As a result, hot-mode accretion is often feeble and $\sim2$ orders of magnitude less intense compared with cold modes (e.g., \citealt{Gaspari_2013}).
Moreover, the hot accretion modes would develop a negative trend between the SMBH mass and plasma entropy/temperature ($\dot M_{\rm B} \propto K_{\rm x}^{-3/2} \propto T_{\rm x}^{-3/2}$), with hotter halos accreting relatively less gas mass, which is ruled out by the observed SMBH mass versus X-ray halo scaling relations (\citealt{Gaspari_2019}) and by our retrieved positive correlation $L_{\rm R} - T_{\rm x}$, a proxy for $P_{\rm BH} - T_{\rm x}$ (Sec. \ref{scalings}).
Finally, no major duty cycle is expected from this mode. This is in conflict with a large intrinsic $\epsilon$ and with what is usually found in groups and clusters. There AGN feedback needs to rapidly suppress cooling of the hot halo via the AGN kinetic/radio power, i.e. establish frequent and efficient self-regulation \citep[][]{McNamara-Nulsen_2007,Gaspari_2011b,Fabian_2012,Gitti_2012,Prasad_2015,Yang_2016}.

Finally, we note that major mergers are unlikely to represent efficient triggers of the AGN and produce the scaling relations discussed here. The typical timescale of $\sim$ 5-6 Gyr between two major mergers \citep[e.g.,][]{Rodriguez_2015} is too long to provide steady support for the feeding of AGN. This is also consistent with \citet{Sharma_2021}, who recently showed that AGN activity is not enhanced by mergers. Moreover, we do not find substantive evidence for violent mergers in our sample.
Nevertheless, mergers can still play a role over time in terms of supplying and preserving a significant amount of gas on the outskirts of the group halo.

\section{Conclusions}

We have carried out a comparison of the kinematic and optical properties of four different samples of COSMOS spectroscopic galaxy members: BGGs with and without radio emission, and satellites with and without radio emission. Scaling relations for the BGG samples were also investigated. Our results can be summarised as follows:

\begin{itemize}
    \item Out of 70 BGGs, 56 ($\sim$80\%) are classified as ancient infallers, while the same fraction for satellites is only $\sim$42\%. We find that the fraction of ancient infaller among radio BGGs is $\sim$82\%, while the fraction of non-radio BGGs falling into this category is  $\sim$78\%. This suggests that most BGGs, and in particular those hosting radio emission, have been accreted by the group at an early time.
    
    \item We find that radio galaxies with $L_R> 10^{23}$ W Hz$^{-1}$ always lie within 0.2\,$R_{\rm vir}$ from the group centre, which has been defined as the X-ray emission peak. This is consistent with the current view of AGN feedback since the gas cooling out of the hot IGrM can feed the central SMBH, while outer galaxies need to rely on more episodic triggers.
    
    \item Our samples were compared to simulated BGGs from the HORIZON-AGN simulation. The ratio between the galaxy line-of-sight velocity and the group velocity dispersion for real BGGs (both radio and non-radio) shows a broader distribution than simulated galaxies, but still narrower than satellites. Statistical tests suggest that significant differences exist between simulated and real galaxies, indicating that additional physics may be needed to reproduce the true population of BGGs.
    
    \item We find that the stellar mass for radio BGGs is statistically higher than for non-radio BGGs. This, in combination with the correlation between BGG mass and group mass, suggests that it is easier to find radio BGGs in higher-mass groups.
    
    \item We find positive correlations between the 1.4 GHz power of radio BGGs and the main properties of the diffuse X-ray halo/intragroup medium, namely, $L_{\rm R} - T_{\rm x}$ and $L_{\rm R} - L_{\rm x}$, suggesting a link between AGN heating and cooling processes in the gaseous halo.
    
    \item We tested and discussed the two major AGN feeding/feedback scenarios.
    Our finding that galaxies at group centres are often radio galaxies better supports the CCA scenario since the AGN can feed from both the galactic and intragroup halo condensations via a flickering duty cycle. This is more difficult to explain in hot accretion modes (e.g., Bondi or ADAF), which can only tap into the nuclear ($r < 100$ pc) pressure-supported plasma region via continuous accretion.
    Unlike in hot modes, CCA becomes more vigorous in hotter and more luminous/massive halos. Thus, CCA naturally induces positive $L_{\rm R} - T_{\rm x}$ and $L_{\rm R} - L_{\rm x}$ correlations, as found in our samples.
    
\end{itemize}

\section*{Acknowledgements}

We thank the referee for thoughtful comments that have improved the presentation of our results. AF thanks Gary Mamon and Stefano Borgani for insightful discussions.
TP is supported by the BMBF Verbundforschung under grant number 50OR1906.
MB acknowledges support from the Deutsche Forschungsgemeinschaft under Germany's Excellence Strategy - EXC 2121 "Quantum Universe" - 390833306. 
MG acknowledges partial support by NASA Chandra GO8-19104X/GO9-20114X and HST GO-15890.020-A grants. \\

\section*{Data Availability}
The data underlying this article are available upon reasonable request.

\newcommand{\newblock}{}
\bibliographystyle{mnras}
\bibliography{bibliography}

\bsp	
\label{lastpage}
\end{document}